\documentclass[12pt,a4paper]{article}
\usepackage{amsmath,amsfonts,amssymb}
\usepackage{hyperref}

\def\bn{\mathbf{n}}

\def\mC{\mathcal{C}}

\def\mH{\mathcal{H}}

\def\bx{\mathbf{x}}
\def\by{\mathbf{y}}
\def \mD{\mathcal{D}}

\newcommand{\tmD}{\tilde{\mathcal{D}}}

\newcommand{\mG}{\mathcal{G}}

\newcommand{\bT}{\mathbf{T}}

\newcommand{\mL}{\mathcal{L}}

\def\pb #1{\left\{#1\right\}}

\newcommand{\phys}{\mathrm{phys}}

\begin{document}

\begin{titlepage}

\begin{center}
{\Large \bf Mimetic dark matter, ghost instability and a mimetic
tensor-vector-scalar gravity}
\end{center}

\vspace{1.5em}

\begin{center}
Masud Chaichian,$^a$ Josef Kluso\v{n},$^b$
Markku Oksanen,$^a$ and Anca Tureanu$^{a,}$\footnote{Email addresses:
masud.chaichian@helsinki.fi (M. Chaichian), klu@physics.muni.cz (J.
Kluso\v{n}), markku.oksanen@helsinki.fi (M. Oksanen),
anca.tureanu@helsinki.fi (A. Tureanu)}\\
\vspace{1em}$^a$\textit{Department of Physics, University of Helsinki,
P.O. Box 64,\\ FI-00014 Helsinki, Finland}\\
\vspace{.3em} $^b$\textit{Department of Theoretical Physics and
Astrophysics, Faculty of Science,\\
Masaryk University, Kotl\'a\v{r}sk\'a 2, 611 37, Brno, Czech Republic}
\end{center}

\vspace{1.5em}

\begin{abstract}
Recently modified gravitational theories which mimic the behaviour of
dark matter, the so-called ``Mimetic Dark Matter'', have been proposed.
We study the consistency of such theories with respect to the absence of
ghost instability and propose a new tensor-vector-scalar theory of
gravity, which is a generalization of the previous models of mimetic
dark matter with additional desirable features. The original model
proposed by Chamseddine and Mukhanov [JHEP 1311 (2013) 135,
arXiv:1308.5410] is concluded to describe a regular pressureless dust,
presuming that we consider only those configurations where the energy
density of the mimetic dust remains positive under time evolution. For
certain type of configurations the theory can become unstable. Both
alternative modified theories of gravity, which are based on a vector
field (tensor-vector theory) or a vector field and a scalar field
(tensor-vector-scalar theory), are free of ghost instabilities.
\end{abstract}

\end{titlepage}
\newpage

\section{Introduction}
Recently a new interesting model of mimetic dark matter was suggested
in \cite{Chamseddine:2013kea} and was further elaborated in
\cite{Barvinsky:2013mea,Chamseddine:2014vna}. The basic idea is
remarkably simple. The physical metric $g_{\mu\nu}^{\phys}$ is
considered to be a function of a scalar field $\phi$ and a fundamental
metric $g_{\mu\nu}$, where the physical metric is defined as\footnote{We
follow the convention used in \cite{Barvinsky:2013mea} and we also
consider the space-time metric of the signature $(-,+,+,+)$.}
\begin{equation}\label{g^phys}
g_{\mu\nu}^{\phys}=\left(-g^{\alpha\beta}\partial_\alpha\phi
\partial_\beta\phi\right) g_{\mu\nu} .
\end{equation}
The physical metric $g_{\mu\nu}^{\phys}$ is invariant with respect to
the Weyl transformation of the metric $g_{\mu\nu}$,
\begin{equation}\label{defg'}
g'_{\mu\nu}(x)=\Omega^2(x) g_{\mu\nu}(x) .
\end{equation}
Then it was shown in \cite{Chamseddine:2013kea} and in
\cite{Barvinsky:2013mea} that the ordinary Einstein-Hilbert action
constructed using the physical metric $g_{\mu\nu}^{\phys}$ possesses
many interesting properties.  In fact, the model analyzed below is a
conformal extension of Einstein's general theory of relativity. The
local Weyl invariance is ensured by introducing an extra degree of
freedom that as was shown in \cite{Chamseddine:2013kea} has the form of
pressureless perfect fluid that, according to
\cite{Chamseddine:2013kea}, can mimic the behavior of a real cold dark
matter.

Historically, Gunnar Nordstr\"om was the first to construct a
relativistic theory of gravity as a scalar field theory
\cite{Nordstrom:1913adp} whose geometric reformulation
\cite{Einstein:1914adp} was the first metric theory of gravity. The
physical metric of this gravitational theory was defined as a conformal
transformation of the flat Minkowski metric,
$g_{\mu\nu}=\phi^2\eta_{\mu\nu}$, where $\phi$ is the scalar field of
Nordstr\"om's theory. In other words, it was a theory of conformally
flat spacetimes. The structure of the field equation, $R=24\pi GT$,
where $R$ and $T$ are the traces of the Ricci tensor and the
energy-momentum tensor respectively, closely resembled the field
equation of the general theory of relativity formulated by Einstein in
the following year. Intriguingly, the idea of mimetic matter
\cite{Chamseddine:2013kea} is to introduce additional fields in the
conformal factor that relates the physical and auxiliary metrics
(\ref{g^phys}) in such a way that the physical metric remains invariant
under the conformal transformation (\ref{defg'}).

The mimetic dark matter proposal \cite{Chamseddine:2013kea} is very
interesting and certainly deserves further study. In general,
formulation of the theory using $g_{\mu\nu}^{\phys}$ can lead to a
theory with higher order derivatives of $\phi$, which may imply the
emergence of ghosts. In order to answer this question, it is necessary
to obtain the Hamiltonian formulation of given theory, identify all
constraints and perform the counting of all local degrees of freedom. A
preliminary analysis of the ghost issue was performed in
\cite{Barvinsky:2013mea}, where however the gauge fixing condition
$g^{\mu\nu}\partial_\mu\phi \partial_\nu\phi+1=0$ was imposed before
proceeding to the canonical formulation. We rather start with the
original action and perform its canonical analysis in full generality.
First we rewrite the action into a form that does not contain
derivatives higher than the first order. Surprisingly we find that the
action resembles the Lagrange multiplier modified action
\cite{Lim:2010yk} (see also \cite{Gao:2010gj,Capozziello:2010uv})
whose Hamiltonian analysis was performed in \cite{Kluson:2010af}.
By solving the second class constraints we derive the Hamiltonian for
the scalar field, which turns out to be linear in the momentum conjugate
to the scalar field. Since this kind of Hamiltonians often are unstable,
especially in higher derivative theories, a careful analysis is
required. The Hamiltonian of the original model
\cite{Chamseddine:2013kea} is argued to be unbounded from below for
certain type of initial configurations and consequently it can become
unstable. Then we perform the Hamiltonian analysis of the Proca vector
field model suggested in \cite{Barvinsky:2013mea}. Since now there are
no derivatives higher than the first order, the Hamiltonian constraint
is found to depend quadratically on the momenta conjugate to the vector
field so that it is bounded from below. This is a very interesting
result that implies that the Proca model should be studied further.
Finally, we present a mimetic tensor-vector-scalar gravity which is a
generalization of the previous models
\cite{Chamseddine:2013kea,Barvinsky:2013mea,Chamseddine:2014vna}
featuring mimetic matter, and that shares some properties with the
celebrated tensor-vector-scalar theories of gravity proposed by
Bekenstein \cite{Bekenstein:2004ne} and Moffat \cite{Moffat:2005si}. It
has been recently shown that Bekenstein's tensor-vector-scalar gravity
is free of ghost degrees of freedom provided that its scalar and vector
fields satisfy a certain condition \cite{Chaichian:2014dfa}.

The organization of this paper is as follows. In section~\ref{second},
we introduce the mimetic dark matter action that was suggested in
\cite{Chamseddine:2013kea}. Then we perform its canonical analysis. We
identify all constraints and determine the number of physical degrees of
freedom. Gauge fixing of the conformal symmetry and the dust structure
of the Hamiltonian are discussed in sections~\ref{Gfs} and
\ref{Dust}. In section~\ref{third}, we discuss the formulation of
the theory in the Einstein frame. In section~\ref{fourth}, we perform
the canonical analysis of the Proca model introduced in
\cite{Barvinsky:2013mea}. In section~\ref{fifth}, we propose a
tensor-vector-scalar theory that is a generalization of the
aforementioned models of mimetic dark energy.

\section{Hamiltonian analysis of mimetic dark matter
model}\label{second} In this section we perform the Hamiltonian
analysis of the mimetic dark matter model that was introduced in
\cite{Chamseddine:2013kea}. The gravitational action is defined as
\begin{equation}\label{Sfund}
S[g_{\mu\nu},\phi]=\frac{1}{2}\int d^4x
\sqrt{-g^{\phys}(g_{\mu\nu},\phi)}
R\left( g_{\mu\nu}^{\phys}(g_{\mu\nu},\phi)\right) ,
\end{equation}
where we have set $8\pi G=1$ and where the physical metric is
parameterized in terms of the fundamental metric $g_{\mu\nu}$ and the
space-time gradients of the scalar field $\phi$ as
\begin{equation}
g_{\mu\nu}^{\phys}=\left(-g^{\alpha\beta}\partial_\alpha\phi
\partial_\beta\phi\right) g_{\mu\nu}\equiv \Phi^2g_{\mu\nu} .
\end{equation}
Matter fields couple to the physical metric minimally. We omit matter
fields in our analysis since their contribution is similar as in general
relativity. To proceed further we express the action using the metric
$g_{\mu\nu}$ rather than the physical metric $g_{\mu\nu}^{\phys}$. This
can be done using the well known relation, see for example
\cite{Faraoni:1998qx},
\begin{equation}\label{Rgphys}
R(g_{\mu\nu}^{\phys})=\frac{1}{\Phi^2}\left(R(g_{\mu\nu})
-6\frac{g^{\mu\nu}\nabla_\mu \nabla_\nu\Phi}{\Phi}\right) ,
\end{equation}
where the covariant derivative $\nabla_\mu$ is defined using the
metric $g_{\mu\nu}$. Inserting (\ref{Rgphys}) into (\ref{Sfund}) we
obtain
\begin{equation}\label{actgPhi}
S[g_{\mu\nu},\phi]
%=\frac{1}{2}\int d^4x \Phi^4\sqrt{-g}[\frac{1}{\Phi^2}
%[R(g)-6\frac{g^{\mu\nu}\nabla_\mu\nabla_\nu\Phi}{\Phi}]] \nonumber\\
% =\frac{1}{2}\int d^4x[\sqrt{-g}\Phi^2R(g_{\mu\nu})
% -6g^{\mu\nu}\Phi
% \nabla_\mu\nabla_\nu\Phi +\Phi^4] \nonumber \\
=\frac{1}{2}\int d^4x\sqrt{-g}\left[ \Phi^2R(g_{\mu\nu})+
6g^{\mu\nu}\nabla_\mu\Phi\nabla_\nu\Phi \right] .
\end{equation}
Clearly the action is invariant under the conformal transformation
(\ref{defg'}) of the metric $g_{\mu\nu}$, since the metric
$g_{\mu\nu}^{\phys}$ is invariant by construction. On the other hand we
see that this action contains second order derivative of $\phi$ so that
we should worry about possible existence of the ghosts. In order to
obtain an action with the first order derivatives, we introduce an
auxiliary field $\lambda$ and rewrite the action (\ref{actgPhi}) into
the form
\begin{equation}\label{Slambda}
S[g_{\mu\nu},\Phi,\lambda,\phi]
%=\frac{1}{2}\int d^4x[ \sqrt{-g} \Phi^2R[-g]+ 6\sqrt{-g}
%g^{\mu\nu}\nabla_\mu\Phi\nabla_\nu\Phi] \nonumber \\
=\frac{1}{2}\int d^4x\sqrt{-g}\left[R(g_{\mu\nu})\Phi^2+6
g^{\mu\nu}\nabla_\mu\Phi\nabla_\nu\Phi -\lambda(
\Phi^2+g^{\mu\nu}\nabla_\mu\phi\nabla_\nu\phi) \right] ,
\end{equation}
where now we treat $\Phi$ as an independent field together with
$\phi$. Note that solving the equation of motion for $\lambda$ we
find $\Phi^2=-g^{\mu\nu}\nabla_\mu\phi\nabla_\nu\phi $. Then
inserting this result into (\ref{Slambda}) we can derive the original
action (\ref{actgPhi}).

Now we can proceed to the Hamiltonian formulation of the theory. We
use the following $3+1$ decomposition of the metric $g_{\mu\nu}$
\cite{Arnowitt:1962hi,Gourgoulhon:2007ue},
\begin{eqnarray}\label{hgdef}
g_{00}=-N^2+N_i h^{ij}N_j , \quad g_{0i}=N_i , \quad
g_{ij}=h_{ij} ,\nonumber \\
g^{00}=-\frac{1}{N^2} , \quad g^{0i}=\frac{N^i}{N^2} ,
\quad g^{ij}=h^{ij}-\frac{N^i N^j}{N^2} ,
\end{eqnarray}
where we have defined $h^{ij}$ as the inverse to the induced metric
$h_{ij}$ on the Cauchy surface $\Sigma_t$ at each time $t$,
\begin{equation}
h_{ik}h^{kj}=\delta_i^{ \ j},
\end{equation}
and we denote $N^i=h^{ij}N_j$. The four dimensional scalar curvature in
$3+1$ formalism has the form
\begin{equation}
R(g_{\mu\nu})=
K_{ij}\mG^{ijkl}K_{kl}+R+\frac{2}{\sqrt{-g}}\partial_\mu(\sqrt{-g}
n^\mu K) -\frac{2}{\sqrt{h}N}\partial_i(\sqrt{h}h^{ij}\partial_j N) ,
\end{equation}
where the extrinsic curvature of the spatial hypersurface $\Sigma_t$ at
time $t$ is defined as
\begin{equation}
K_{ij}=\frac{1}{2N}\left(\frac{\partial h_{ij}}{\partial t}- D_i
N_j-D_j N_i\right) ,
\end{equation}
with $D_i$ being the covariant derivative determined by the metric
$h_{ij}$, and where the de Witt metric is defined as
\begin{equation}
\mG^{ijkl}=\frac{1}{2}(h^{ik}h^{jl}+h^{il}h^{jk}) -h^{ij}h^{kl}
\end{equation}
with inverse
\begin{equation}
\mG_{ijkl}=\frac{1}{2}(h_{ik}h_{jl}+h_{il}h_{jk}) -\frac{1}{2}
h_{ij}h_{kl} \
\end{equation}
that obeys the relation
\begin{equation}
\mG_{ijkl}\mG^{klmn}=\frac{1}{2}(\delta_i^m\delta_j^n+\delta_i^n
\delta_j^m) .
\end{equation}
Further, $n^\mu$ is the future-pointing unit normal vector to the
hypersurface $\Sigma_t$, which is written in terms of the ADM variables
as
\begin{equation}
n^0=\sqrt{-g^{00}}=\frac{1}{N} ,\quad
n^i=-\frac{g^{0i}}{\sqrt{-g^{00}}}=-\frac{N^i}{N} .
\end{equation}
Inserting these results to the action and performing integration by
parts we obtain the action in the form
\begin{eqnarray}\label{actHam}
S[N,N^i,h_{ij},\Phi,\lambda,\phi]&=&\frac{1}{2} \int dt d^3\bx\sqrt{h}N
\left[ K_{ij}\mG^{ijkl}K_{kl}\Phi^2 +R\Phi^2
-4K\Phi\nabla_n\Phi \right.\nonumber\\
&-&\frac{2}{\sqrt{h}}\partial_i(\sqrt{h}h^{ij}\partial_j\Phi^2)
-6(\nabla_n\Phi)^2+6h^{ij}\partial_i\Phi\partial_j\Phi
\nonumber\\
&-&\left.\lambda\Phi^2 +\lambda(\nabla_n\phi)^2
-\lambda h^{ij}\partial_i\phi\partial_j\phi \right] ,
\end{eqnarray}
where
\begin{equation}
\nabla_n\Phi=\frac{1}{N}(\partial_t\Phi-N^i\partial_i\Phi) ,
\end{equation}
and where we ignored boundary terms.  Now we can easily derive
the momenta conjugate to $h_{ij},\Phi,\lambda$ and $\phi$ from the
action (\ref{actHam}) as
\begin{eqnarray}
\pi^{ij}&=&\frac{1}{2}\sqrt{g}\mG^{ijkl}K_{kl}\Phi^2
-\sqrt{h}h^{ij}\nabla_n\Phi\Phi , \nonumber \\
p_\Phi&=&-2K\Phi\sqrt{h}-6\sqrt{h}\nabla_n\Phi , \nonumber \\
p_\lambda&\approx & 0 , \quad  p_\phi=\sqrt{h}\lambda\nabla_n\phi .
\end{eqnarray}
Using these relations we obtain the following primary constraint
\begin{equation}
\mD=p_\Phi\Phi-2\pi^{ij}h_{ij}\approx 0
\end{equation}
and the Hamiltonian in the form
\begin{eqnarray}\label{defHT}
H&=&\int d^3\bx \left( N\mH_T+N^i\mH_i+v_\mD \mD+v_N\pi_N
+v^i\pi_i+v_\lambda p_\lambda \right) , \nonumber \\
\mH_T&=& \frac{2}{\sqrt{h}\Phi^2}\pi^{ij}\mG_{ijkl}\pi^{kl}-
\frac{1}{2}\sqrt{h}R\Phi^2 +\frac{1}{2\sqrt{h}\lambda}p_\phi^2
+\partial_i(\sqrt{h}h^{ij}\partial_j\Phi^2) \nonumber \\
&-&3\sqrt{h}h^{ij}\partial_i\Phi\partial_j\Phi
+\frac{1}{2}\sqrt{h}\lambda\left( \Phi^2 +h^{ij}\partial_i\phi
\partial_j\phi\right) , \nonumber \\
\mH_i&=&p_\Phi\partial_i\Phi+p_\phi\partial_i\phi
-2h_{ij}D_k\pi^{jk} .
\end{eqnarray}
Note that we have ignored the boundary contribution of the Hamiltonian,
since we are interested in the behavior of the local degrees of freedom,
rather than in the total gravitational energy. Therefore the Hamiltonian
\eqref{defHT} is a sum of constraints that vanishes for any physical
configuration on the constraint surface. In addition to the constraint
terms, a complete Hamiltonian contains a surface term on the boundary
of space, what defines the total energy of the system.
The total energy is conserved in time, and according to the positive
energy theorem of general relativity \cite{Schoen:1979zz,Witten:1981cmp}
the total energy is positive, except for flat Minkowski spacetime, which
has zero energy.\footnote{The nonminimally coupled field $\Phi$ is not
dynamical, since it is a gauge degree of freedom associated with the
conformal symmetry. The total gravitational energy is independent of the
chosen gauge for the conformal symmetry. When we fix the gauge of the
conformal symmetry in section~\ref{Gfs}, we obtain a minimally coupled
scalar field theory, which is known to describe dust (see
sections~\ref{Dust} and \ref{third}). We require that the energy density
of the scalar field is positive on the initial Cauchy surface, say at
time $t=0$, since only those initial configurations are physically
meaningful. Then the energy conditions of the positive energy theorem of
general relativity are satisfied at the inital time $t=0$ and the total
gravitational energy is positive. Since the total energy is conserved,
it remains positive. We will later argue that the system can become
unstable when the energy density of the scalar field becomes negative
under time evolution. In that case, the gravitational field should
compensate for the contribution of the scalar field so that the total
energy remains conserved.}
Our analysis concerns the structure of the Hamiltonian density and the
properties of the local degrees of freedom, and their constraints.

Now we proceed to the analysis of the preservation of the primary
constraints $\pi_N\approx 0 , \pi_i\approx 0,\mD\approx 0 ,
p_\lambda\approx 0$. As usual the requirement of the preservation of
the constraints $\pi_N,\pi_i$ implies the secondary constraints
\begin{equation}\label{mH=0}
\mH_T\approx 0 ,\quad \mH_i\approx 0 .
\end{equation}
For further analysis we introduce the smeared form of these
constraints
\begin{equation}
\bT_T(N)=\int d^3\bx N\mH_T ,\quad \bT_S(N^i)= \int d^3\bx
(N^i\mH_i+p_\lambda \partial_i\lambda) .
\end{equation}
On the other hand the requirement of the preservation of the
constraint $p_\lambda$ implies
\begin{eqnarray}
\frac{1}{N}\partial_t p_\lambda=\frac{1}{N}\pb{p_\lambda,H}=
\frac{1}{2\sqrt{h}\lambda^2}p_\phi^2 -\frac{1}{2}\sqrt{h}\left(
\Phi^2 +h^{ij}\partial_i\phi\partial_j\phi \right)
\equiv \mC_\lambda\approx 0 .
\end{eqnarray}
Let us now proceed to the requirement of the preservation of the
constraint $\mD$. However, it is convenient to consider the following
linear combination of $\mD$ with the constraint $p_\lambda\approx 0$
in the form
\begin{equation}\label{tmD}
\tmD=\mD+2p_\lambda\lambda= p_\Phi\Phi-2\pi^{ij} h_{ij}+ 2p_\lambda
\lambda,
\end{equation}
which has the following non-zero Poisson brackets:
\begin{eqnarray}
\pb{\tmD(\bx),h_{ij}(\by)}&=& 2h_{ij}(\bx)\delta(\bx-\by) ,
\nonumber\\
\pb{\tmD(\bx),\pi^{ij}(\by)}&=&-2\pi^{ij}(\bx)\delta(\bx-\by) ,
\nonumber \\
\pb{\tmD(\bx),\Phi(\by)}&=&-\Phi(\bx)\delta(\bx-\by) , \nonumber \\
\pb{\tmD(\bx),p_\Phi(\by)}&=& p_\Phi(\bx)\delta(\bx-\by) , \nonumber
\\
\pb{\tmD(\bx),\lambda(\by)}&=&-2\lambda(\bx)\delta(\bx-\by) ,
\nonumber \\
\pb{\tmD(\bx),p_\lambda(\by)}&=& 2p_\lambda(\bx)\delta(\bx-\by) .
\end{eqnarray}
Then we obtain
\begin{eqnarray}
\pb{\tmD,\bT_T(N)}
%-\frac{2N}{\sqrt{h}\Phi^2}\pi^{ij}\mG_{ijkl}\pi^{kl}
%-\frac{1}{2}[-N\sqrt{h}R-4\nabla^i\nabla_i[N\Phi^2]\sqrt{h}]
% \nonumber\\
%+\frac{N}{2\sqrt{h}\lambda}p_\phi^2+\partial_i N
%h^{ij}\partial_j\Phi^2\sqrt{h}-
%2\partial_i[\partial_j Nh^{ij}\sqrt{h}\Phi^2] \nonumber \\
%+3N\sqrt{h}h^{ij}\partial_i\Phi\partial_j\Phi -6\partial_i[
%N\sqrt{h}h^{ij}\partial_j\Phi \Phi] \nonumber\\
%+N\frac{1}{2}\sqrt{h}\lambda h^{ij}\partial_i\phi\partial_j\phi
%+N\frac{1}{2}\sqrt{h}\lambda\Phi^2 \nonumber \\
=-N\mH_T,
\end{eqnarray}
using
\begin{equation}
\delta R=D^iD^j\delta h_{ij}-h^{ij}D^kD_k\delta
h_{ij}-R^{ij}\delta h_{ij} .
\end{equation}
In the same way we find
\begin{eqnarray}
\pb{\tmD,\bT_S(N^i)}=
%\partial_k(N^k p_\Phi\Phi)
%-2\partial_k (N^k \pi^{ij}h_{ij})=
\partial_i(N^i\tmD) .
\end{eqnarray}
Collecting all these results together we find
\begin{eqnarray}
\partial_t\tmD=\pb{\tmD,H_T}=
-N\mH_T+\partial_i(N^i\tmD)+2v_\lambda p_\lambda\approx 0,
\end{eqnarray}
so that $\tmD$ is preserved without imposing any additional
constraint.

Finally we determine Poisson brackets between $\mH_T$ and $\mH_i$.
We use their smeared form and we find
\begin{eqnarray}
\pb{\bT_T(N),\bT_T(M)}&=&
%\int d^3\bx \frac{2}{\sqrt{h}\Phi^2}
%\pi^{ij}\mG_{ijkl}(N\nabla^k\nabla^l(\sqrt{h}\Phi^2M)-
%N\nabla^m\nabla_m (h^{kl}\Phi^2M) \nonumber \\
%-M\nabla^k\nabla^l(\sqrt{h}\Phi^2N)- M\nabla^m\nabla_m
%(h^{kl}\Phi^2N) \nonumber \\
% +\int d^3\bx (N\partial_jM-M\partial_jN)h^{ji}
%p_\phi\partial_i\phi
%\nonumber \\
%+\int d^3\bx(-(N\partial_mM-M\partial_iN)
%\frac{2\partial_n\Phi^2}{\sqrt{g}\Phi^2}
%\pb{\pi^{ij}\mG_{ijkl}\pi^{kl}(\bx),\sqrt{h}h^{mn}(\by)} \nonumber
%\\
% =\int d^3\bx (\partial_mMN-\partial_mNM)
%(-2\nabla_n\pi^{ij}\mG_{ijkl}h^{lm}h^{nk}+2 h^{mn}\nabla_n\pi^{ij}
%\mG_{ijkl}h^{kl}) \nonumber \\
%+8\int d^3\bx \pi^{ij}(\delta_i^m\delta_j^n)
%(\partial_mMN-\partial_mNM)\frac{\partial_n\Phi}{\Phi} \nonumber
%\\
% +\int d^3\bx (N\partial_jM-M\partial_jN)h^{ji}
%p_\phi\partial_i\phi  \nonumber \\
%-8\int d^3\bx (N\partial_mM-M\partial_nN)\frac{\partial_n\Phi}{\Phi}
%(\delta_i^n\delta_l^m-\frac{1}{4}h_{ij}h^{mn})\pi^{ij}
%\nonumber \\
%=\int d^3\bx(N\partial_iM-M\partial_iN) h^{ji} (-2h_{ik}\nabla_m
%\pi^{mk}+p_\phi\partial_i\phi) \nonumber \\
%+2\int d^3\bx(N\partial_mN-M\partial_m N)\frac{\partial_n\Phi}{
%\Phi}\pi^{kl}h_{kl}g^{mn} \nonumber \\
\bT_S((N\partial_i M-M\partial_iN)h^{ij})  \nonumber \\
&-&\int
d^3\bx(\partial_iMN-N\partial_iM)h^{ij}\frac{\partial_j\Phi}{\Phi}\mD
\end{eqnarray}
and we see that given expression vanishes on the constraint surface
$\mH_i\approx  0, \mD\approx 0$. Further we have
\begin{eqnarray}
\pb{\bT_S(N^i),\bT_S(M^i)}=\bT_S((N^i\partial_iM^j-M^i\partial_iN^j))
\end{eqnarray}
and finally
\begin{eqnarray}
\pb{\bT_S(N^i),\bT_T(M)}=\bT_T(N^i\partial_iM) .
\end{eqnarray}
Hence the Hamiltonian and momentum constraints (\ref{mH=0}) are
preserved under time evolution.

The number of physical degrees of freedom is obtained via Dirac's
formula: (number of canonical variables)/2 $-$ (number of first class
constraints) $-$ (number of second class constraints)/2. Compared to
general relativity, we have six extra canonical variables
($\phi,p_\phi,\Phi,p_\Phi,\lambda,p_\lambda$), one extra first class
constraint $\tmD\approx0$, and two extra second class constraints
$p_\lambda\approx0$, $\mC_\lambda\approx0$. Thus, in addition to the two
gravitational degrees of freedom of general relativity, there exist one
extra physical degree of freedom.

Now we see that $p_\lambda\approx0$ and $\mC_\lambda\approx0$ are the
second class constraints that can be set to vanish strongly. We solve
the constraint $\mC_\lambda=0$ with respect to $\lambda$ as
\begin{equation}\label{lambdadef}
\lambda=\pm\frac{p_\phi}{\sqrt{h}\sqrt{\Phi^2+h^{ij}
\partial_i\phi\partial_j\phi}}.
\end{equation}
Inserting each of these two solutions into the Hamiltonian constraint
defined in (\ref{defHT}), we find that it is equal to
\begin{eqnarray}\label{mHTfinal}
\mH_T&=& \frac{2}{\sqrt{h}\Phi^2}\pi^{ij}\mG_{ijkl}\pi^{kl}-
\frac{1}{2}\sqrt{h}R\Phi^2+\partial_i
(\sqrt{h}h^{ij}\partial_j\Phi^2) \nonumber \\
&-&3\sqrt{h}h^{ij}\partial_i\Phi\partial_j\Phi
\pm p_\phi\sqrt{\Phi^2+h^{ij}\partial_i\phi\partial_j\phi}.
\end{eqnarray}
% using
% \begin{equation}
% \frac{1}{2\sqrt{h}\lambda}p_\phi^2 +\frac{1}{2}\sqrt{h}\lambda
% \left( h^{ij}\partial_i\phi\partial_j\phi+\Phi^2 \right)=
% \pm p_\phi\sqrt{\Phi^2+h^{ij}\partial_i\phi\partial_j\phi}.
% \end{equation}
The pair of canonical variables $\lambda,p_\lambda$ has been eliminated
from the formalism. Then we should determine the Dirac bracket between
the remaining phase space variables $\pi^{ij},h_{ij},\phi,p_\phi$ and
$\Phi,p_\Phi$. Recall that by definition the Dirac bracket between two
phase space functions has the form
\begin{equation}
\pb{A,B}_D=\pb{A,B}-\sum_{I,J}\pb{A,\Psi_I}(\Omega^{-1})^{IJ}
\pb{\Psi_J,B} ,
\end{equation}
where $\Psi^I, I=1,2$ are the second class constraints
$p_\lambda\approx 0, \mC_\lambda\approx 0$ and where $\Omega_{IJ}$ is
the matrix of the Poisson brackets
\begin{eqnarray}
\Omega_{p_\lambda(\bx),\mC_\lambda(\by)}&=&\pb{p_\lambda(\bx),
\mC_\lambda(\by)}=\frac{1}{\sqrt{h}\lambda^3}
p_\phi^2(\bx)\delta(\bx-\by)
,  \nonumber \\
\Omega_{\mC_{\lambda}(\bx),\mC_\lambda(\by)}&=&
\frac{1}{\sqrt{h}\lambda^2(\bx)}p_\phi(\bx)\sqrt{h}h^{ij}(\by)
\partial_{y^j}\delta(\bx-\by)\partial_{y^i}\phi(\by)
\nonumber \\
&-&\frac{1}{\sqrt{h}\lambda^2(\by)}p_\phi(\by)\sqrt{h}h^{ij}(\bx)
\partial_{x^j}\delta(\bx-\by)\partial_{x^i}\phi(\bx),
\end{eqnarray}
so that the matrix $\Omega$ has the following schematic form
\begin{equation}
\Omega=\left(\begin{array}{cc} 0 & A \\
-A & B \\
\end{array}\right)
\end{equation}
and its inverse has the form
\begin{equation}
\Omega^{-1}=\left(\begin{array}{cc} A^{-1}BA^{-1} & -A^{-1} \\
A^{-1} & 0 \\ \end{array}\right).
\end{equation}
Now we find that the Dirac brackets coincides with the Poisson brackets
due to the fact that $\pb{h_{ij},p_\lambda}=
\pb{\pi^{ij},p_\lambda}=\pb{p_\phi,p_\lambda}=\pb{\phi,p_\lambda}=0$.

Note that the scalar part of the Hamiltonian has a similar structure as
in the case of Lagrange multiplier modified gravities, see
\cite{Kluson:2010af}. The reason for this becomes evident in
section~\ref{third}, where we find that by gauge fixing the conformal
symmetry we can write the theory as a certain type of Lagrange
multiplier modified gravity.

Let us now return to the Hamiltonian with the second class constraints
solved. From (\ref{mHTfinal}) we see that the Hamiltonian constraint
depends linearly on the momentum $p_\phi$. In some cases such a linear
dependence implies that the Hamiltonian density is not bounded from
below, which is a classical sign of instability. This is especially the
case in many higher derivative field theories, which are notorious for
their Ostrogradskian instability (for discussion and example, see
\cite{Woodard:2007,Kluson:2013hza}). On the other hand, in certain cases
a linear dependence on a canonical momentum does not imply instability.
As a very simple example we mention the Hamiltonian of a harmonic
oscillator written in terms of action-angle variables, $H=\omega P$,
where $\omega$ is the angular frequency and $P$ is the momentum
conjugate to the action-angle coordinate (see, e.g.,
\cite{Chaichian:2012mec}). Since a momentum that is conjugate to an
action-angle coordinate is a constant of motion, the previous
Hamiltonian is a constant and hence trivially bounded from below. Thus,
in order to determine whether the present theory of mimetic dark matter
\cite{Chamseddine:2013kea} involves an instability, we have to study the
dynamics of the variables $\phi,p_\phi$ carefully. In the present case,
the stability of the system depends on whether the momentum $\pm p_\phi$
in the two alternative Hamiltonians given by (\ref{mHTfinal}) can evolve
to the negative side of the phase space, $\pm p_\phi<0$, and eventually
to negative infinity, $\pm p_\phi\rightarrow-\infty$. If that happens,
the system is unstable. Note that when the kinetic $p_\phi$-term
becomes negative, the metric/tensor part of the Hamiltonian constraint
$\mathcal{H}_T$ has to compensate for it by increasing its local value,
since the Hamiltonian constraint must remain zero for every physical
configuration. Consequently, the system could be driven to an
increasingly excited state, one kinetic term towards negative infinity
and another term towards positive infinity, and hence a stable vacuum
could not exist.

We will later show that the two alternative Hamiltonians given by
(\ref{mHTfinal}) actually describe the same physical system. Therefore
it suffices to consider the dynamics for one of the cases, which we
choose to be the one with the linear dependence on $+p_\phi$. Namely,
we consider the Hamiltonian given by
\begin{eqnarray}\label{mHT+}
\mH_T&=& \frac{2}{\sqrt{h}\Phi^2}\pi^{ij}\mG_{ijkl}\pi^{kl}-
\frac{1}{2}\sqrt{h}R\Phi^2+\partial_i
(\sqrt{h}h^{ij}\partial_j\Phi^2) \nonumber \\
&-&3\sqrt{h}h^{ij}\partial_i\Phi\partial_j\Phi
+p_\phi\sqrt{\Phi^2+h^{ij}\partial_i\phi\partial_j\phi}.
\end{eqnarray}
Physically, the momentum $p_\phi$ is proportional to the energy density
of the mimetic dust on the spatial hypersurface $\Sigma_t$. More
precisely, $p_\phi$ is the rest mass density of the mimetic dust per
coordinate volume element $d^3x$, as measured by the Eulerian observers
with four-velocity $n^\mu$. Since $p_\phi$ has the physical meaning of
density of rest mass, we require that $p_\phi$ is initially nonnegative
everywhere. That is the initial configuration of the system must satisfy
$p_\phi\ge0$ everywhere on the initial Cauchy surface, say $\Sigma_0$
at time $t=0$. The physical meaning of $\phi$ is that its gradient
$\partial_\mu\phi$ is the direction of the rest mass current of the
mimetic dust in spacetime. Then let us discuss the dynamics of the
system and in particular the dynamics of the mimetic dust.
We obtain the equation of motion for $\phi$ in the form
\begin{equation}\label{p_tphi}
\partial_t\phi=\pb{\phi,H}=N\sqrt{\Phi^2+h^{ij}
\partial_i\phi\partial_j\phi}+N^i\partial_i\phi .
\end{equation}
The square of this equation gives the relation of $\Phi$ to $\phi$ as
$\Phi^2=(\nabla_n\phi)^2-h^{ij}\partial_i\phi\partial_j\phi
=-g^{\mu\nu}\partial_\mu\phi\partial_\nu\phi$, which tells that the rest
mass current of the mimetic dust is a timelike vector in spacetime.
From the equation of motion (\ref{p_tphi}) we obtain that the time
evolution of $\phi$ does not depend on $p_\phi$, which is a rather
strange equation of motion.
This kind of systems where the evolution of a coordinate does not
depend on canonical momenta have been studied in the past in the context
of 't Hooft's deterministic quantum mechanics
\cite{'tHooft:2001ct,'tHooft:1999gk,Blasone:2009yp,Blasone:2004yf}.
Presuming a gauge where $N=\text{positive constant}$, $N^i=0$ and
$\Phi=\text{positive constant}$, we see from (\ref{p_tphi}) that
$\phi$ experiences monotonic and eternal growth under time evolution.
The rate of growth has the minimal value of $\partial_t\phi=N\Phi$ and
it is speed up by the presence of spatial nonhomogeneity in $\phi$ so
that the norm of the spacetime gradient $\partial_\mu\phi$ remains
constant. The spatial gradient of $\phi$ is the dynamically relevant
quantity, while the local value of $\phi$ on $\Sigma_t$ is physically
irrelevant.

The equation of motion for the momentum $p_\phi$ has the form
\begin{equation}\label{p_tp_phi}
\partial_tp_\phi=\pb{p_\phi,H}=\partial_i\left(\frac{Np_\phi
h^{ij}\partial_j\phi}{\sqrt{\Phi^2+h^{ij}\partial_i\phi
\partial_j\phi}}+N^ip_\phi\right) .
\end{equation}
Physically this is the continuity equation for the rest mass current of
the mimetic dust, which ensures that the total rest mass on the spatial
hypersurface $\Sigma_t$ is conserved under time evolution from one
hypersurface to the next. First we would like to find the configuration
of this system which could be interpreted as the ground state in the
sense that the time derivative of $p_\phi$ is equal to zero. We can
rewrite the equation of motion
(\ref{p_tp_phi})  as
\begin{equation}\label{p_tp_phi.2}
\partial_tp_\phi=p_\phi\partial_i\left(\frac{Nh^{ij}\partial_j\phi}
{\sqrt{\Phi^2+h^{ij}\partial_i\phi\partial_j\phi}}+N^i\right)
+\partial_ip_\phi\left(\frac{Nh^{ij}\partial_j\phi}
{\sqrt{\Phi^2+h^{ij}\partial_i\phi
\partial_j\phi}}+N^i\right) .
\end{equation}
From (\ref{p_tp_phi.2}) we see that there exists a ground state where
$p_\phi=0$. If there exists a region of space where $p_\phi=0$, then
inside that region $p_\phi$ remains zero, since $p_\phi=0$ and
$\partial_ip_\phi=0$ imply that $\partial_tp_\phi=0$. Then let us
consider an initial configuration where $p_\phi>0$ (inside some region
or everywhere in space), which corresponds to the presence of mimetic
dust. Now the crucial question is whether $p_\phi$ can evolve to the
negative side of the phase space $p_\phi<0$. If that can happen, the
system is unstable. Indeed, since negative $p_\phi$ means dust with
negative rest mass, an infinite amount of radiation, matter or dust
could be created without violating the conservation of total energy. The
question has two steps: can $p_\phi$ evolve to zero, and if it does, can
it become negative? Assuming the aforementioned gauge, we obtain the
equation of motion (\ref{p_tp_phi.2}) as
\begin{equation}\label{p_tp_phi.gf}
\partial_tp_\phi=Np_\phi\partial_i\left(\frac{h^{ij}\partial_j\phi}
{\sqrt{\Phi^2+h^{ij}\partial_i\phi\partial_j\phi}}\right)
+\frac{Nh^{ij}\partial_ip_\phi\partial_j\phi}
{\sqrt{\Phi^2+h^{ij}\partial_i\phi\partial_j\phi}} .
\end{equation}
Since the evolution of $\phi$ (and of its gradient $\partial_i\phi$)
with time is independent of $p_\phi$, we can setup the configuration for
$\phi$ freely when we consider the dynamics of $p_\phi$.
Consider a (small) region of space where the metric $h_{ij}$ and the
gradient $\partial_i\phi$ are nearly constant. Then the equation of
motion (\ref{p_tp_phi.gf}) is dominated by its second term, while the
first term is very small in comparison.  Furthermore, we consider
that the gradient $\partial_i\phi$ is contradirectional compared to the
gradient $\partial_ip_\phi$ so that
$h^{ij}\partial_ip_\phi\partial_j\phi<0$.  For example, let us consider
that the given point is a local minimum of $p_\phi$, so that
$\partial_ip_\phi$ is pointing away from the given point. Thus,
regardless of how close $p_\phi$ is to zero, it can evolve towards zero,
since $\partial_tp_\phi$ can be negative. There appears to be nothing
that could stop $p_\phi$ from evolving to zero, since the time evolution
of $p_\phi$ does not necessarily change the direction of the gradient
$\partial_ip_\phi$ so that $h^{ij}\partial_ip_\phi\partial_j\phi$ would
become nonnegative. However, proving this decisively would require an
exact solution that crosses the point $p_\phi=0$. Since such a solution
must be nonhomogeneous and nonisotropic, it is very hard to achieve.
Alternatively, one could try to show that on some background the
perturbation of $p_\phi$ can grow to negative infinity.
Let us then consider what happens assuming that $p_\phi$ has evolved to
zero. Now the equation of motion (\ref{p_tp_phi.gf}) reads as
\begin{equation}\label{p_tp_phi.zero.gf}
\partial_tp_\phi=\frac{Nh^{ij}\partial_ip_\phi\partial_j\phi}
{\sqrt{\Phi^2+h^{ij}\partial_i\phi\partial_j\phi}} .
\end{equation}
When the directions of the gradients of $p_\phi$ and $\phi$ are such
that $h^{ij}\partial_ip_\phi\partial_j\phi<0$, we obtain that
$\partial_tp_\phi<0$ and consequently $p_\phi$ becomes negative.
Thus, our arguments indicate that under certain circumstances, the
energy density of the mimetic dust can become negative, and consequently
the system can become unstable.

Let us consider the Hamiltonian constraint (\ref{mHTfinal}) with
the negative sign in front of $p_\phi$. Now $-p_\phi$ has the physical
meaning as the rest mass density of the mimetic dust. Hence $p_\phi$
must be negative initially. The equations of motion are obtained as
\begin{equation}
\partial_t\phi=\pb{\phi,H}=-N\sqrt{\Phi^2+h^{ij}
\partial_i\phi\partial_j\phi}+N^i\partial_i\phi
\end{equation}
and
\begin{equation}
\partial_tp_\phi=\pb{p_\phi,H}=\partial_i\left(-\frac{Np_\phi
h^{ij}\partial_j\phi}{\sqrt{\Phi^2+h^{ij}\partial_i\phi
\partial_j\phi}}+N^ip_\phi\right) .
\end{equation}
We can see that this system is simply the mirror image of the system
considered above obtained via the transformation
$(\phi,p_\phi)\rightarrow(-\phi,-p_\phi)$.

We briefly consider the inclusion of a potential term for the scalar
field $\phi$. It would be included into the Hamiltonian constraint
$\mH_T$ by adding $\sqrt{h}\Phi^4 V(\phi)$, where the potential
$V(\phi)$ is a local function of $\phi$. Then it would contribute an
extra term into the right-hand side of the equation of motion
(\ref{p_tp_phi}) as $-N\sqrt{h}\Phi^4\frac{dV(\phi)}{d\phi}$. When this
term is positive, it raises the bar for the appearance of the
instability, but this does not change the conclusion. The system can
still become unstable for the given kind of initial configurations.

Despite the potential problem of instability discussed above the
original theory of mimetic dark matter could be useful for astrophysical
and cosmological modeling, provided that one considers only
those initial configurations that do not cross the point $p_\phi=0$
under time evolution. Those are the cases that describe physical dust.

\subsection{Gauge fixing of scale symmetry}\label{Gfs}
Returning to our Hamiltonian formulation we fix the dilatation symmetry
by introducing the gauge fixing function
\begin{equation}\label{Phi=1}
\mC_{\mathrm{scale}}\equiv\Phi-1=0 .
\end{equation}
Clearly we have
\begin{equation}
\pb{\mC_{\mathrm{scale}}(\bx),\mD(\by)}=\delta(\bx-\by)
\end{equation}
and hence they are the second class constraints. We can also
explicitly solve $\mD$ for $p_\Phi$ and we obtain
\begin{equation}
p_\Phi=2\pi^{ij}h_{ij} .
\end{equation}
Then the Hamiltonian constraint has the form
\begin{equation}\label{mHTgf}
\mH_T=\mH_T^{GR}+\mH_T^\phi \approx0,
\end{equation}
where $\mH_T^{GR}$ is the standard contribution of general relativity,
\begin{equation}
\mH_T^{GR}=\frac{2}{\sqrt{h}}\pi^{ij}\mG_{ijkl}\pi^{kl}-
\frac{1}{2}\sqrt{h}R,
\end{equation}
and the contribution of the scalar field is given as
\begin{equation}
\mH_T^\phi=\frac{1}{2\sqrt{h}\lambda}p_\phi^2
+\frac{1}{2}\sqrt{h}\lambda\left( 1+h^{ij}\partial_i\phi
\partial_j\phi\right).
\end{equation}
The momentum constraint is given as
\begin{equation}
 \mH_i=p_\phi\partial_i\phi-2h_{ij}D_k\pi^{jk}\approx 0.
\end{equation}
Solving the second class constraints $p_\lambda\approx0$ and
\begin{equation}
\mC_\lambda=\frac{1}{2\sqrt{h}\lambda^2}p_\phi^2
+\frac{1}{2}\sqrt{h}\left( 1+h^{ij}\partial_i\phi\partial_j\phi \right)
\approx0,
\end{equation}
we again find the contribution of the scalar $\phi$ in the form
\begin{equation}\label{mHTphi}
\mH_T^\phi=p_\phi\sqrt{1+h^{ij}\partial_i\phi\partial_j\phi},
\end{equation}
where we chose the solution (\ref{lambdadef}) with the positive sign.
Now only the first class constraints which are associated with the
diffeomorphism invariance remain, namely $\pi_N\approx0$,
$\pi_i\approx0$, $\mH_T\approx0$, $\mH_i\approx0$.
There exists an extra scalar degree of freedom that is associated with
the variables $\phi$, $p_\phi$. It couples to the metric via the square
root factor in (\ref{mHTphi}). The term that is linear in the momentum
$p_\phi$ persists and it has the same form and dynamics which were
discussed above.

\subsection{Dust structure of the Hamiltonian}\label{Dust}
If we assume that $\lambda>0$ everywhere in spacetime, and consequently
that $p_\phi>0$, we obtain the contribution of the scalar field
(\ref{mHTphi}) in the Hamiltonian (\ref{mHTgf}) as
\begin{equation}\label{mHTphidust}
\mH_T^\phi=\sqrt{p_\phi^2+h^{ij}\mH_i^\phi\mH_j^\phi},
\end{equation}
where $\mH_i^\phi$ denotes the contribution of $\phi$ to the momentum
constraint in (\ref{defHT}),
\begin{equation}
\mH_i^\phi=p_\phi\partial_i\phi.
\end{equation}
The same Hamiltonian for dust was obtained in \cite{Brown:1994py} using
eight scalar fields on spacetime to describe the full dynamics of dust.
Once the conformal gauge is fixed (see section ~\ref{Gfs}) the mimetic
theory contains only two extra scalars, namely the field $\phi$ and the
solvable Lagrange multiplier $\lambda$. In \cite{Brown:1994py}, these
two scalars are denoted by $T$ and $M$, respectively. $M$ was assumed to
be positive since it represents the rest mass density of the dust.
Compared to \cite{Brown:1994py} the present mimetic model lacks both the
dust frame fields $Z^k$ ($k=1,2,3$) and the spatial components $W_k$ of
the four-velocity of dust in the dust frame. Thus we conclude that the
mimetic theory \cite{Chamseddine:2013kea} with the assumption
$\lambda>0$, and with the conformal symmetry gauge fixed, is a reduced
version of the model of dust that was studied in \cite{Brown:1994py}.

The problem with imposing the condition $\lambda>0$ is that it appears
to be inconsistent with the equations of motion in some cases. When
$\lambda$ is solved (\ref{lambdadef}), the requirement $\lambda>0$
becomes the requirement $p_\phi>0$ (or $p_\phi<0$ for the second
solution). Since $p_\phi$ is the density of the mimetic dust, we require
that $p_\phi>0$ on the initial Cauchy surface. The arguments presented
above indicate that for some initial configurations, $p_\phi$ could
evolve to zero and further to $p_\phi<0$. Hence the requirement
$\lambda>0$ is not always consistent with the dynamics. Actually, the
requirement $\lambda>0$ appears to be equivalent to picking up the
initial configurations that do no cross the surface $p_\phi=0$ under
time evolution.

\section{Mimetric theory in Einstein frame}\label{third}
In this section, we present the formulation of the mimetric theory in
the Einstein frame. We can rewrite the action (\ref{Slambda}) in the
Einstein frame by gauge fixing the dilatation symmetry. When we perform
the conformal transformation $g_{\mu\nu}\rightarrow\Omega^2g_{\mu\nu}$,
using
\begin{equation}
R(g_{\mu\nu})\rightarrow\frac{1}{\Omega^2}\left( R(g_{\mu\nu})
-6\frac{g^{\mu\nu}\nabla_\mu\nabla_\nu\Omega}{\Omega} \right),
\end{equation}
and with the scale fixed as $\Omega=\Phi^{-1}$,  the action
(\ref{Slambda}) takes the following form
\begin{equation}\label{SgfE}
S[g_{\mu\nu},\lambda,\phi]=\frac{1}{2}\int d^4x\sqrt{-g}
\left[ R(g_{\mu\nu}) -\lambda\left( 1+g^{\mu\nu}
\nabla_\mu\phi\nabla_\nu\phi \right) \right],
\end{equation}
where $\lambda$ is the Lagrange multiplier that ensures the spacetime
gradient $\nabla_\mu\phi$ is unit and timelike.
The same action can be obtained by simply setting the conformal gauge in
(\ref{Slambda}) as $\Phi=1$. Thus the conformal invariance is
necessarily lost in the Einstein frame. This reduced form of the mimetic
theory was first obtained in \cite{Golovnev:2013jxa} but in another way.
Now we see that in the Einstein frame the mimetic theory reduces to
Lagrange multiplier modified gravity \cite{Lim:2010yk} without any
extra terms present in the Lagrangian except the constraint multiplied
by $\lambda$. The Hamiltonian analysis of Lagrange multiplier modified
theory of gravity was performed in \cite{Kluson:2010af} with additional
kinetic and potential terms included.
The Hamiltonian for the action (\ref{SgfE}) is the same one that was
obtained in section \ref{Gfs} via gauge fixing the conformal symmetry
of the mimetic theory with (\ref{Phi=1}). In other words, the
formulation in Einstein frame represents one conformal gauge of the
mimetic theory. As usual, there exist alternative gauges.

\section{Vector field model of mimetic dark matter}\label{fourth}
The vector field model of mimetic dark matter theory was suggested in
\cite{Barvinsky:2013mea}. It is based on the presumption that the
physical metric has the form
\begin{equation}
g_{\mu\nu}^{\phys}=-(g^{\alpha\beta}u_\alpha
u_\beta)g_{\mu\nu}\equiv \Phi^2 g_{\mu\nu}.
\end{equation}
The action that now contains the Maxwell kinetic term has the form
\begin{equation}\label{vectoraction}
S[g_{\mu\nu}^{\phys},u_\mu]= \int d^4x \sqrt{-g^{\phys}}
\left[\frac{1}{2}R(g_{\mu\nu}^{\phys})-\frac{\mu^2}{4}g^{\mu\alpha}_{
\phys}
g^{\nu\beta}_{\phys}F_{\mu\nu}F_{\alpha\beta}\right] ,
\end{equation}
where
\begin{equation}
g^{\mu\nu}_{\phys}=\Phi^{-2}g^{\mu\nu}
\end{equation}
and
\begin{equation}
F_{\mu\nu}=\partial_\mu u_\nu-\partial_\nu u_\mu ,
\end{equation}
and where $\mu^2$ is the parameter having the dimension of mass squared.

Our goal is to perform the Hamiltonian analysis of given action. We
follow the analysis performed in the case of scalar action so that we
obtain\footnote{In principle there is no compelling reason why to
introduce $\Phi$ as an independent variable. However, the presence of
the term $g^{\mu\nu}\nabla_\mu\Phi\nabla_\nu\Phi$ in the action would
imply the following expression $ \Phi^{-2}g^{\mu\nu}(\nabla_\mu u_\rho
g^{\rho\sigma}u_\sigma)(\nabla_\nu u_\gamma
g^{\gamma\delta}u_\delta)$ that would lead to very complicated
expression when we implemented $3+1$ decomposition of the metric and
the vector field $u_\mu$. For that reason we still treat $\Phi$ as
an independent variable exactly as in the previous section.}
\begin{eqnarray}
S[g_{\mu\nu},\Phi,\lambda,u_\mu]=\int d^4x \sqrt{-g}
\left[\frac{1}{2}\Phi^2R(g_{\mu\nu})
+3g^{\mu\nu}\nabla_\mu\Phi\nabla_\nu\Phi \right.\nonumber \\
+\left.\frac{1}{2} \lambda(\Phi^2+g^{\mu\nu}u_\mu
u_\nu)-\frac{\mu^2}{4}g^{\mu\alpha}g^{\nu\beta}
F_{\mu\nu} F_{\alpha\beta} \right].
\end{eqnarray}
Now we are ready to proceed to the Hamiltonian formalism of given
action. The gravitational part is the same as before. On the other
hand in case of the vector part we closely follow
\cite{Chaichian:2014dfa} and find that the action for the vector
field has the form
\begin{eqnarray}\label{SAADM}
S_{u}&=&-\frac{\mu^2}{4}\int d^4x\sqrt{-g}
g^{\mu\alpha}g^{\nu\beta}F_{\mu\nu} F_{\alpha\beta}
\nonumber \\
&=&-\frac{\mu^2}{4} \int dt d^3\bx \sqrt{h}N
[h^{ik}h^{jl}(D_iu_j-D_ju_i)(D_ku_l-D_lu_k) \nonumber \\
 &-& 2h^{ij}(
\mL_n u_i -a_i u_\bn-D_iu_\bn)(\mL_n u_j-a_ju_\bn-D_ju_\bn)] ,
\end{eqnarray}
where \begin{equation} \mL_n u_i= \frac{1}{N}(\partial_t
u_i-\mL_{\vec{N}}u_i)= \frac{1}{N}(\partial_t u_i-N^k\partial_k
u_i-\partial_iN^ku_k)
 , a_i=\frac{D_iN}{N} \ .
\end{equation}
Using this form of the  action we can easily proceed to the
Hamiltonian formulation. First of all we have following conjugate
momenta $p^i,p_\bn$ to $u_i$ and $u_\bn$, respectively
\begin{equation}
p^i=\frac{\delta L}{\delta \partial_t u_i}= \mu^2\sqrt{h}h^{ij}
(\mL_n u_j-a_j u_{\bn}-D_ju_{\bn}) ,\quad
 p_{\bn}\approx 0 .
\end{equation}
Then it is easy to find the Hamiltonian for given system
\begin{eqnarray}
H&=&\int d^3\bx \left( N\mH_T+N^i\mH_i+v_\mD \mD+v_N\pi_N
+v^i\pi_i+v_\lambda p_\lambda+v_\bn p_\bn \right) , \nonumber \\
\mH_T&=& \frac{2}{\sqrt{h}\Phi^2}\pi^{ij}\mG_{ijkl}\pi^{kl}-
\frac{1}{2}\sqrt{h}R\Phi^2 +\partial_i
[\sqrt{h}h^{ij}\partial_j\Phi^2] \nonumber \\
&-&3\sqrt{h}h^{ij}\partial_i\Phi
\partial_j\Phi-\frac{1}{2}\sqrt{h}\lambda(\Phi^2
+u_i h^{ij}u_j-u_\bn^2) \nonumber \\
&+&\frac{1}{2\mu^2\sqrt{h}} p^ih_{ij}p^j +\frac{\mu^2}{4}\sqrt{h}
h^{ik}h^{jl}(D_iu_j-D_ju_i)(D_ku_l-D_lu_k)-u_\bn D_i p^i
 , \nonumber \\
\mH_i&=&p_\Phi\partial_i\Phi-2h_{ik}D_j \pi^{kj} +\partial_i u_j
p^j-\partial_j (u_i p^j) .
\end{eqnarray}
Note that we have
\begin{equation}
\pb{\bT_S(N^i),u_j(\bx)}=-N^k\partial_k u_i-\partial_i N^k u_k \ .
\end{equation}
Now we can proceed with the analysis in the same way as in previous
section. We have the following set of the primary
constraints\footnote{Note that it is not surprising that we include the
expression $p_{\bn}u_{\bn}$ into the definition of the dilatation
constraint, since it can be easily shown that $u_{\bn}$ transforms
non-trivially under (\ref{defg'}). In fact, by definition we have
$u_{\bn}=n^\mu u_\mu=\frac{1}{N}(u_0-N^iu_i)$ and since $N,N^i$
transform under (\ref{defg'}) as $N'=\Omega N , N'_i=\Omega^2 N_i,
N'^i=N^i$, we easily find that $u'_\bn=-\frac{1}{\Omega}u_\bn$.}
\begin{eqnarray}
\tmD=p_\Phi\Phi-2\pi^{ij}h_{ij}+2p_\lambda\lambda+p_{\bn}u_{\bn}\approx
0 , \nonumber \\
p_\lambda\approx 0 , \quad  \pi_N\approx 0 , \quad  \pi_i \approx
0 , \quad  p_{\bn}\approx 0 .
\end{eqnarray}
The preservation of the primary constraints $p_N,p^i$ during the
time evolution of the system implies following secondary constraints
\begin{equation}
\mH_T\approx 0 ,\quad \mH_i\approx 0 ,
\end{equation}
while the constraint $\tmD$ is preserved and it is the first class
constraint that is the generator of the scaling transformation. On the
other hand the requirement of the preservation of the constraint
$p_\lambda\approx 0$ implies
\begin{eqnarray}
\frac{1}{N}\partial_t p_\lambda=\frac{1}{N}\pb{p_\lambda, H}
=\frac{1}{2}\sqrt{h}\left(\Phi^2+h^{ij}u_i u_j-u_{\bn}^2\right)
\equiv\frac{1}{2}\mC_\lambda\approx 0.
\end{eqnarray}
Finally the requirement of the preservation of the constraint
$p_{\bn}$ gives
\begin{eqnarray}
\frac{1}{N}\partial_t p_{\bn}=\frac{1}{N}\pb{p_{\bn},H}
=-\sqrt{h}\lambda u_{\bn}+D_i p^i\equiv \mC_\bn\approx 0 .
\end{eqnarray}
To proceed further note that we have following non-zero Poisson
brackets
\begin{eqnarray}
\pb{p_\lambda(\bx), \mC_{\bn}(\by)}=\sqrt{h}u_\bn(\bx)\delta(\bx-\by)
, \nonumber \\
\pb{p_{\bn}(\bx),\mC_{\bn}(\by)}=\sqrt{h}\lambda(\bx)\delta(\bx-\by)
 , \nonumber \\
\pb{p_{\bn}(\bx),\mC_\lambda(\by)}=2\sqrt{h}u_{\bn}(\bx)\delta(\bx-\by),
\end{eqnarray}
which show that $p_\lambda,\mC_{\bn},p_{\bn}$ and $\mC_{\lambda}$ are
the second class constraints.
Considering the Hamiltonian constraint, we find after some
calculations
\begin{eqnarray}
\pb{\bT_T(N),\bT_T(M)}&=& \bT_S((N\partial_i M-M\partial_i N)h^{ij})
\nonumber \\
&-&\int d^3\bx (\partial_i MN-N\partial_i
M)h^{ij}\frac{\partial_j\Phi}{\Phi}\mD \nonumber \\
&-&\int d^3\bx (\partial_i MN-M\partial_i M)h^{ij}u_j\mC_{\bn}.
\end{eqnarray}
This again implies that $\mH_T$ and $\mH_i$ are the first class
constraints which is the reflection of the diffeomorphism invariance
of given theory.

Let us now return to the second class constraints in given model.
 We saw that $\mC_{\bn}$ and $\mC_{\lambda}$ are
the second class constraints that vanish strongly. From
$\mC_{\lambda}$ we can express $u_{\bn}$ as
\begin{equation}\label{ubnsol}
u_{\bn}=-\sqrt{\Phi^2+u_i h^{ij}u_j} ,
\end{equation}
where we chose $u_{\bn}$ to be negative for a reason that will be clear
below. The sign of $u_{\bn}$ can be chosen since $u_{\bn}$ is no longer
an independent variable. Further, from $\mC_{\bn}$ we express $\lambda$
as
\begin{equation}
\lambda=-\frac{D_ip^i}{\sqrt{h} \sqrt{\Phi^2+u_i h^{ij}u_j}} .
\end{equation}
Using these results we find the Hamiltonian constraint in the form
\begin{eqnarray}\label{mhTvectfin}
\mH_T&=& \frac{2}{\sqrt{h}\Phi^2}\pi^{ij}\mG_{ijkl}\pi^{kl}-
\frac{1}{2}\sqrt{h}R\Phi^2 +\partial_i
[\sqrt{h}h^{ij}\partial_j\Phi^2] \nonumber \\
&-&3\sqrt{h}h^{ij}\partial_i\Phi
\partial_j\Phi+\frac{1}{2\mu^2\sqrt{h}} p^ih_{ij}p^j
+\sqrt{\Phi^2+u_i h^{ij}u_j}D_ip^i \nonumber
\\
&+&\frac{\mu^2}{4}\sqrt{h}
h^{ik}h^{jl}(D_iu_j-D_ju_i)(D_ku_l-D_lu_k) .
\end{eqnarray}
There exist five physical degrees of freedom, since we have sixteen
pairs of canonical variables, nine first class constraints
($\mH_T,\mH_i,\pi_N,\pi_i,\mD$) and four second class constraints
($p_\lambda,p_\bn,\mC_\lambda,\mC_\bn$).
We can again gauge fix the scale symmetry as in section~\ref{Gfs} in
order to obtain the Hamiltonian constraint in the form
\begin{equation}
\mH_T=\mH_T^{\mathrm{GR}}+\mH_T^u,
\end{equation}
where the contribution of the vector field is given as
\begin{eqnarray}\label{mhTvectgf}
\mH_T^u&=&\frac{1}{2\mu^2\sqrt{h}} p^ih_{ij}p^j
+\sqrt{1+u_i h^{ij}u_j}D_ip^i \nonumber\\
&+&\frac{\mu^2}{4}\sqrt{h}h^{ik}h^{jl}(D_iu_j-D_ju_i)
(D_ku_l-D_lu_k) .
\end{eqnarray}
From (\ref{mhTvectfin}) or (\ref{mhTvectgf}) we can read following
important information. Three physical degrees of freedom are carried in
the vector field and two in the metric.  The Hamiltonian constraint
$\mH_T$ depends on the momenta $p^i$ quadratically and due to the fact
that we have chosen $-$ sign in front of the square root (\ref{ubnsol})
we also see that the vector part is positive definite and hence bounded
from below. In other words there is no sign of the ghost instability.
In summary, the vector form of the mimetic model seems to be very
promising model of the dark energy and deserves to be elaborated
further.

\section{Mimetic tensor-vector-scalar gravity}\label{fifth}
We propose a tensor-vector-scalar theory of gravity that is a
generalization of the theories of mimetic dark matter
\cite{Chamseddine:2013kea,Barvinsky:2013mea}.
The gravitational action of this theory includes both a vector field
$u_\mu$ and a scalar field $\phi$. Several different
tensor-vector-scalar theories of gravity have been proposed in order to
address the problem of dark matter, most notably the theories proposed
by Bekenstein \cite{Bekenstein:2004ne} and Moffat \cite{Moffat:2005si}.
Our proposal differs from all the previous tensor-vector-scalar
theories.

The physical metric is defined to have the form
\begin{equation}\label{g.tvs}
g_{\mu\nu}^{\phys}=-f(\phi)(g^{\alpha\beta}u_\alpha u_\beta)
g_{\mu\nu}\equiv \Phi^2 g_{\mu\nu},
\end{equation}
where $f(\phi)$ is some (dimensionless) nonnegative function of $\phi$,
e.g., $f(\phi)\propto\phi^2$. The physical metric is again invariant
under the conformal transformation of the metric $g_{\mu\nu}$.
The standard kinetic term and an optional potential term for the scalar
field $\phi$ are included into the action (\ref{vectoraction}) so that
the total gravitational action has the form
\begin{eqnarray}
S[g_{\mu\nu}^{\phys},u_\mu,\phi]= \int d^4x \sqrt{-g^{\phys}}
\left[\frac{1}{2}R(g_{\mu\nu}^{\phys})
-\frac{\mu^2}{4}g^{\mu\alpha}_{\phys}g^{\nu\beta}_{\phys}
F_{\mu\nu}F_{\alpha\beta} \right. \nonumber\\
-\left.\frac{1}{2}g^{\mu\nu}_{\phys}\nabla_\mu^{\phys}\phi
\nabla_\nu^{\phys}\phi -V(\phi)
+L(g_{\mu\nu}^{\phys},\chi,\partial\chi,u_\mu,\phi) \right] .
\end{eqnarray}
The Lagrangian $L(g_{\mu\nu}^{\phys},\chi,\partial\chi,u_\mu,\phi)$
represents matter and its interaction with the gravitational fields.
Variation of the action with respect to the physical metric
$g_{\mu\nu}^{\phys}$ implies the standard gravitational field equations.
The variation is given as
\begin{eqnarray}
\delta_{g_{\mu\nu}^{\phys}} S
% &=&\frac{1}{2}\int d^4x\sqrt{-g^{\phys}}
% \left( -G^{\mu\nu}_{\phys} +T_F^{\mu\nu} +T_\phi^{\mu\nu}
% +T_{\phys}^{\mu\nu} \right)\delta g_{\mu\nu}^{\phys} \nonumber\\
% &+&\frac{1}{2}\int d^4x\sqrt{-g^{\phys}}\left(
% \nabla^\mu_{\phys}\nabla^\nu_{\phys}\delta g_{\mu\nu}^{\phys}
% -g^{\mu\nu}_{\phys}\nabla^2_{\phys} \delta g_{\mu\nu}^{\phys} \right)
%  \nonumber\\
&=&\frac{1}{2}\int d^4x\sqrt{-g^{\phys}}
\left( -G^{\mu\nu}_{\phys} +T_F^{\mu\nu} +T_\phi^{\mu\nu}
+T^{\mu\nu} \right)\delta g_{\mu\nu}^{\phys} \nonumber\\
&-&\oint d^3x\sqrt{|\gamma^{\phys}|}\delta K^{\phys},
\label{deltaS}
\end{eqnarray}
where
\begin{equation}
G^{\mu\nu}_{\phys}=R^{\mu\nu}(g_{\mu\nu}^{\phys})
-\frac{1}{2}g^{\mu\nu}_{\phys}R(g_{\mu\nu}^{\phys}),
\end{equation}
\begin{equation}
T_F^{\mu\nu}=\mu^2\left( F^{\mu\alpha}F^\nu_{\phantom\nu\alpha}
-\frac{1}{4}g^{\mu\nu}_{\phys}F_{\alpha\beta}F^{\alpha\beta} \right),
\end{equation}
\begin{equation}
T_\phi^{\mu\nu}=\nabla^\mu_{\phys}\phi \nabla^\nu_{\phys}\phi
-g^{\mu\nu}_{\phys}\left( \frac{1}{2}\nabla_\alpha^{\phys}\phi
\nabla^\alpha_{\phys}\phi +V(\phi) \right)
\end{equation}
and
\begin{equation}
T_{\phys}^{\mu\nu}=\frac{2}{\sqrt{-g^{\phys}}}
\frac{\delta}{\delta g_{\mu\nu}^{\phys}}\int d^4x\sqrt{-g^{\phys}}
L(g_{\mu\nu}^{\phys},\chi,\partial\chi,u_\mu,\phi).
\end{equation}
Note that above everything is written in the physical frame, e.g.,
$F^{\alpha\beta}=g^{\alpha\mu}_{\phys}g^{\beta\nu}_{\phys}F_{\mu\nu}$.
The second integral in (\ref{deltaS}) is taken over the boundary of
spacetime with $\gamma^{\phys}$ and $K^{\phys}$ being the determinant of
the induced metric and the trace of the extrinsic curvature on the
boundary, respectively. This surface contribution could be canceled by
adding an appropriate boundary term into the action.

When the action is viewed as a functional of $g_{\mu\nu}$, $u_\mu$,
$\phi$ and the matter fields, the variation of the physical metric reads
as
\begin{eqnarray}
\delta g_{\mu\nu}^{\phys}(g_{\mu\nu},u_\mu,\phi)
&=&\Phi^2\delta g_{\mu\nu}
+g_{\mu\nu}\delta\Phi^2 \nonumber\\
&=&\Phi^2\delta g_{\mu\nu}+g_{\mu\nu}f(\phi)g^{\alpha\rho}
g^{\alpha\sigma}u_\rho u_\sigma \delta g_{\alpha\beta} \nonumber\\
&-&2g_{\mu\nu}f(\phi)g^{\alpha\beta}u_\alpha \delta u_\beta
-g_{\mu\nu}f'(\phi)(g^{\alpha\beta}u_\alpha u_\beta)\delta\phi
\nonumber\\
&=&\Phi^2\left( \delta_\mu^\alpha \delta_\nu^\beta +g_{\mu\nu}^{\phys}
f(\phi)g^{\alpha\rho}_{\phys}g^{\alpha\sigma}_{\phys}
u_\rho u_\sigma \right)\delta g_{\alpha\beta} \nonumber\\
 &-&2g_{\mu\nu}^{\phys}f(\phi)g^{\alpha\beta}_{\phys}
 u_\alpha \delta u_\beta
-g_{\mu\nu}^{\phys}f'(\phi)(g^{\alpha\beta}_{\phys}u_\alpha
u_\beta)\delta\phi .
\end{eqnarray}
Hence we obtain the following gravitational field equations:
\begin{eqnarray}
G^{\mu\nu}_{\phys}=T_{\phys}^{\mu\nu}+\varepsilon u^\mu u^\nu
+T_F^{\mu\nu} +T_\phi^{\mu\nu},\\
\mu^2\nabla_\nu^{\phys}F^{\nu\mu}-\varepsilon u^\mu
+\frac{\partial L}{\partial u_\mu}=0,\\
\nabla_\mu^{\phys}\nabla^\mu_{\phys}\phi
-\varepsilon u_\mu u^\mu\frac{f'(\phi)}{2f(\phi)}
-V'(\phi) +\frac{\partial L}{\partial\phi}=0,
\end{eqnarray}
where $u^\mu=g^{\mu\nu}_{\phys}u_\nu$ is the velocity of the mimetic
dust,  whose energy density is given as
\begin{equation}\label{dustenergy}
\varepsilon=f(\phi)g_{\mu\nu}^{\phys}\left( T_{\phys}^{\mu\nu}
+T_\phi^{\mu\nu} -G^{\mu\nu}_{\phys} \right).
\end{equation}
The metric $g_{\mu\nu}$ enters the field equations only through the
physical metric (\ref{g.tvs}), while the vector field and the scalar
field appear also explicitly. The main difference compared to the
vector field model \cite{Barvinsky:2013mea} is the presence of the extra
scalar field $\phi$ in the energy of the dust (\ref{dustenergy}). The
fields $\phi,u_\mu,g_{\mu\nu}$ are also coupled to each other in a
rather intricate way, even in the absence of matter fields. These are
the characteristics that make $\phi$ differ from a regular scalar
field that can be included into the Lagrangian of matter.

\subsection{Hamiltonian formulation}
We again introduce $\Phi$ as an independent variable, rewriting the
action as
\begin{eqnarray}
S[g_{\mu\nu},\Phi,\lambda,u_\mu,\phi]=\int d^4x \sqrt{-g}
\left [\frac{1}{2}\Phi^2R(g_{\mu\nu})
+3g^{\mu\nu}\nabla_\mu\Phi\nabla_\nu\Phi \right.\nonumber \\
+\frac{1}{2} \lambda\left( \Phi^2+f(\phi)g^{\mu\nu}u_\mu
u_\nu \right)
-\frac{\mu^2}{4}g^{\mu\alpha}g^{\nu\beta}
F_{\mu\nu} F_{\alpha\beta} \nonumber\\
-\left.\frac{1}{2}\Phi^2 g^{\mu\nu}\nabla_\mu\phi\nabla_\nu\phi
-\Phi^4 V(\phi) \right].
\end{eqnarray}
We again ignore the boundary terms since we are interested in the local
(propagating) degrees of freedom, rather than the invariant surface
energy.

In the ADM description, the scalar field action has the form
\begin{equation}
S_\phi=\frac{1}{2}\int dtd^3x \sqrt{h}N \Phi^2 \left[ (\nabla_n\phi)^2
-h^{ij}\partial_i\phi\partial_j\phi -\Phi^4 V(\phi) \right],
\end{equation}
where
\begin{equation}
\nabla_n\phi=\frac{1}{N}\left( \partial_t\phi-N^i\partial_i\phi \right).
\end{equation}
The canonical momentum conjugate to $\phi$ is defined as
\begin{equation}
p_\phi=\sqrt{h}\Phi^2 \nabla_n\phi.
\end{equation}
The other canonical momenta --- as well as all the primary constraints
--- are the same as for the vector field model in section~\ref{fourth}.
We obtain the Hamiltonian as
\begin{eqnarray}
H&=&\int d^3\bx \left( N\mH_T+N^i\mH_i+v_\mD \mD+v_N\pi_N
+v^i\pi_i+v_\lambda p_\lambda+v_\bn p_\bn \right) , \nonumber \\
\mH_T&=& \frac{2}{\sqrt{h}\Phi^2}\pi^{ij}\mG_{ijkl}\pi^{kl}-
\frac{1}{2}\sqrt{h}R\Phi^2 +\partial_i
[\sqrt{h}h^{ij}\partial_j\Phi^2] \nonumber \\
&-&3\sqrt{h}h^{ij}\partial_i\Phi
\partial_j\Phi-\frac{1}{2}\sqrt{h}\lambda\left(\Phi^2
+f(\phi)u_i h^{ij}u_j-f(\phi)u_\bn^2\right) \nonumber \\
&+&\frac{1}{2\mu^2\sqrt{h}} p^ih_{ij}p^j +\frac{\mu^2}{4}\sqrt{h}
h^{ik}h^{jl}(D_iu_j-D_ju_i)(D_ku_l-D_lu_k)-u_\bn D_i p^i \nonumber \\
&+&\frac{p_\phi^2}{2\sqrt{h}\Phi^2}
+\frac{1}{2}\sqrt{h}\Phi^2 h^{ij}\partial_i\phi \partial_j\phi
+\Phi^4 V(\phi)
, \nonumber \\
\mH_i&=&p_\Phi\partial_i\Phi-2h_{ik}D_j \pi^{kj} +\partial_i u_j
p^j-\partial_j (u_i p^j)+p_\phi\partial_i\phi .
\end{eqnarray}
The secondary constraints that ensure the preservation of
$p_\lambda\approx0$ and $p_\bn\approx0$ are defined as
\begin{equation}
\mC_\lambda=\sqrt{h}\left( \Phi^2+f(\phi)u_i h^{ij}u_j
-f(\phi)u_\bn^2  \right)\approx0
\end{equation}
and
\begin{equation}
\mC_\bn=-\sqrt{h}\lambda f(\phi)u_\bn+D_ip^i\approx0,
\end{equation}
respectively. These are second-class constraints. The constraint
$\mC_\lambda=0$ can be solved for
\begin{equation}
u_\bn=-\sqrt{\frac{\Phi^2}{f(\phi)}+u_i h^{ij}u_j},
\end{equation}
assuming $f(\phi)>0$. Then $\lambda$ is fixed by the constraint
$\mC_\bn=0$.
% \begin{equation}
% \lambda=\frac{D_ip^i}{\sqrt{h}f(\phi)u_\bn}
% =-\frac{D_ip^i}{\sqrt{h}\sqrt{f(\phi)\Phi^2+f^2(\phi)u_i h^{ij}u_j}}
% \end{equation}
The Hamiltonian constraint is given as
\begin{eqnarray}
\mH_T&=& \frac{2}{\sqrt{h}\Phi^2}\pi^{ij}\mG_{ijkl}\pi^{kl}-
\frac{1}{2}\sqrt{h}R\Phi^2 +\partial_i
[\sqrt{h}h^{ij}\partial_j\Phi^2] \nonumber \\
&-&3\sqrt{h}h^{ij}\partial_i\Phi
\partial_j\Phi \nonumber \\
&+&\frac{1}{2\mu^2\sqrt{h}} p^ih_{ij}p^j
+\sqrt{\frac{\Phi^2}{f(\phi)}+u_i h^{ij}u_j} D_i p^i \nonumber \\
&+&\frac{\mu^2}{4}\sqrt{h}
h^{ik}h^{jl}(D_iu_j-D_ju_i)(D_ku_l-D_lu_k) \nonumber \\
&+&\frac{p_\phi^2}{2\sqrt{h}\Phi^2}
+\frac{1}{2}\sqrt{h}\Phi^2 h^{ij}\partial_i\phi \partial_j\phi
+\Phi^4 V(\phi) .
\label{H_T.tvs}
\end{eqnarray}
There exist six physical degrees of freedom, since we have seventeen
pairs of canonical variables, nine first class constraints
($\mH_T,\mH_i,\pi_N,\pi_i,\mD$) and four second class constraints
($p_\lambda,p_\bn,\mC_\lambda,\mC_\bn$).
Furthermore, one can fix the conformal gauge as in (\ref{Phi=1}),
obtaining a true tensor-vector-scalar theory without auxiliary fields,
whose Hamiltonian constraint is given as
\begin{equation}\label{H_T.tvs.gf}
\mH_T=\mH_T^{\mathrm{GR}}+\mH_T^{u,\phi},
\end{equation}
where the contribution of the vector and scalar fields is given as
\begin{eqnarray}
\mH_T^{u,\phi}&=&\frac{1}{2\mu^2\sqrt{h}} p^ih_{ij}p^j
+\sqrt{\frac{1}{f(\phi)}+u_i h^{ij}u_j} D_i p^i \nonumber \\
&+&\frac{\mu^2}{4}\sqrt{h}
h^{ik}h^{jl}(D_iu_j-D_ju_i)(D_ku_l-D_lu_k) \nonumber \\
&+&\frac{p_\phi^2}{2\sqrt{h}}
+\frac{1}{2}\sqrt{h}h^{ij}\partial_i\phi \partial_j\phi+V(\phi) .
\end{eqnarray}
Alternatively, one could fix the conformal gauge, for example, as
$\Phi^2=f(\phi)$, i.e., in a similar way as in the vector field model,
$-g^{\mu\nu}u_\mu u_\nu=1$.

The Hamiltonian (\ref{H_T.tvs}) or (\ref{H_T.tvs.gf}) has quite similar
characteristics compared to the vector case in section~\ref{fourth}.
Three physical degrees of freedom are carried in the vector field, one
in the scalar field $\phi$, and two in the metric. All the fields have
well-defined kinetic terms. Presuming the potential $V(\phi)$ is stable,
the addition of the extra field $\phi$ does not imply any evident
problems. The scalar mode $\phi$ is coupled to both the tensor and
vector modes, as well as to the scalar $\Phi$.

\section{Conclusions}
We have studied the potential existence of ghost instability in the
recently proposed gravitational theories featuring mimetic dark matter.
The original mimetic dark matter model \cite{Chamseddine:2013kea} ---
that is a conformal extension of Einstein's gravity containing an extra
scalar field that can mimic dusty dark matter --- obtains a Hamiltonian
that is linear in the momentum $p_\phi$ conjugate to the scalar field
$\phi$. The Lagrange multiplier $\lambda$ describes the rest mass
density of the mimetic dust, which is required to be positive on the
initial Cauchy surface. It is solved in terms of $p_\phi$ and the other
canonical variables in (\ref{lambdadef}). If we require that $\lambda>0$
at all times, the Hamiltonian has a structure that is known to describe
dust \cite{Brown:1994py}. However, for certain type of initial
configurations, the requirement $\lambda>0$ is inconsistent with the
dynamics governed by the equations of motion. We indeed argue that
there exist configurations for which the momentum $p_\phi$ can evolve to
zero and then become negative. In those cases, the density of the
mimetic dust becomes negative under time evolution, which means that the
system becomes unstable. Actually, the requirement $\lambda>0$
appears to be equivalent to choosing the initial configurations that
do no cross the surface $p_\phi=0$ under time evolution. Only those
configurations can be used to describe physical systems. Lastly, in
section~\ref{third}, the formulation of the theory in the Einstein frame
is achieved by gauge fixing the conformal symmetry of the original
action.

The alternative model with an extra vector field
\cite{Barvinsky:2013mea} is shown to be well defined from the
Hamiltonian point of view. This model was shown in
\cite{Barvinsky:2013mea} to be able to mimic both potential and
rotational flows of a pressureless perfect fluid. Finally, we have
presented a mimetic tensor-vector-scalar theory of gravity which
contains both a scalar field and a vector field. It is shown to possess
a healthy canonical structure that is free of ghost degrees of freedom.
The inclusion of the scalar field further generalizes the dynamics,
letting both the scalar and vector degrees of freedom contribute to the
mimetic matter.

Phenomenological implications of the tensor-vector-scalar gravity
proposed in section~\ref{fifth} should be studied carefully. The
inclusion of a scalar field in addition to a vector field certainly has
a significant effect compared to the previous models
\cite{Chamseddine:2013kea,Barvinsky:2013mea,Chamseddine:2014vna}.
Implications to cosmology and structure formation are among the main
interests. The scalar field $\phi$ could play a role similar to a
conventional inflaton field. In that case, does its unusual coupling to
$\Phi$ imply any advantage over a conventional minimally coupled
inflaton field, or is the inclusion of $\phi$ into (\ref{g.tvs}) just an
unnecessary complication? Likewise, for a different potential and
couplings, the scalar field $\phi$ could produce the late-time
acceleration, mimicking dark energy.

\subsection*{Acknowledgements}
We thank Andrei Barvinsky and Viatcheslav Mukhanov for several
clarifying discussions. The support of the Academy of Finland under the
Project Nos. 136539 and 272919, as well as of the Magnus Ehrnrooth
Foundation, is gratefully acknowledged. The work of J.K. was supported
by the Grant Agency of the Czech Republic under the grant P201/12/G028.
The work of M.O. was supported by the Jenny and Antti Wihuri Foundation.

\end{document}